\documentclass{aa}
\usepackage[varg]{txfonts}

\usepackage{amsmath}
\usepackage{amssymb}

\usepackage[T1]{fontenc}

\usepackage[english]{babel}
\usepackage{longtable}
\usepackage{graphicx}
\usepackage{verbatim}

\usepackage{makecell}
\bibpunct{(}{)}{;}{a}{}{,} 

\begin{document}

\title{Dark Matter Massive Fermions and Einasto Profiles in Galactic Haloes}

\author{I. Siutsou
  \and C. R. Arg\"uelles
  \and R. Ruffini}

\offprints{I. Siutsou, \email{siutsou@icranet.org}}

\institute{International Center of Relativistic Astrophysics Network (ICRANet),
Piazza
della Repubblica 10, 65122 Pescara, Italy.\smallskip\\
Dipartimento di Fisica, Universit\`a degli Studi di Roma ``Sapienza'', P.le
Aldo Moro 5, 00185 Roma, Italy.}

\date{Received  / Accepted }

\abstract{On the basis of a fermionic dark matter model we fit rotation curves
of The HI Nearby Galaxy Survey THINGS sample and compare our 3-parametric model
to other models widely used in the literature: 2-parametric
Navarro--Frenk--White, pseudoisothermal sphere, Burkhert models, and
3-parametric Einasto model, suggested as the new "standard dark matter profile"
model in the paper by Chemin et. al., AJ 142 (2011) 109. The results from the
fitting procedure provides evidence for an underlying fermionic nature of the
dark matter candidate, with rest mass above the keV regime.}

\keywords{Dark matter -- Astroparticle physics -- Galaxies: halos -- Galaxies:
kinematics and dynamics} \maketitle

\section{Introduction}
The problem of the distribution of Dark Matter (DM) in galaxies, as usually
addressed in the literature, is mainly focused in the halo regions and
associated with the galaxy rotation curves obtained from the observations, see
e.g. \cite{2013BrJPh..43..369E}. A well-known approach used to deal with this
problem is the Navarro--Frenk--White (NFW) model \citep{1997ApJ...490..493N},
expected to provide a universal description of dark matter halos obtained under
the following main considerations: $1)$ N-body simulations in Cold dark matter
(CDM) and ($\Lambda$CDM) cosmologies; $2)$ particles each of masses of $\sim
10^9 M_{\odot}$ \footnote{Modern numerical simulations can reach better
resolution down to particle masses of $\sim 10^5 M_{\odot}$
\citep{2012MNRAS.425.2169G}.}; $3)$ classical Newtonian physics.

Despite an indicated agreement of this model with the large scale structure of
the Universe, some problems remains at galactic scales, see e.g.
\citet{2013ApJ...766...56M}. A central characteristic of the NFW dark matter
density profiles, is that they show a cuspy and divergent behaviour through the
center of the configuration, while empirical profiles tend to show a core of
constant density, giving rise to the well-known core-cusp controversy, see e.g.
\citet{2010AdAst2010E...5D}.

Yet another important approach developed to understand the distribution of
matter in galaxies has been advanced by \citet{1965TrAlm...5...87E} and
\citet{1989A&A...223...89E}. This is a phenomenological approach consisting in
the proposal of an empirical fitting function composed by three free parameters
as detailed in equation (\ref{eq:2})
\begin{equation}
\rho_E(r)=\rho_{-2}\exp{\left(-\frac{2}{n}\left[\left(\frac{r}{r_{-2}}\right)^n-1\right]\right)},
\label{eq:2}
\end{equation}
where $\rho_{-2}$ and $r_{-2}$ are the density and radius at which
$\rho(r)\propto r^{-2}$, and $n$ is the Einasto index which determines the
shape of the profile.

Recent N-body simulations in $\Lambda$CDM cosmology by
\citet{2004MNRAS.349.1039N} purported a novel dark matter halo model different
from NFW. This model was soon realized \citep{2006AJ....132.2685M} to be the
same as the Einasto one as given by equation (\ref{eq:2}).

\begin{figure}[!h]
\centerline{\includegraphics[width=0.85\columnwidth]{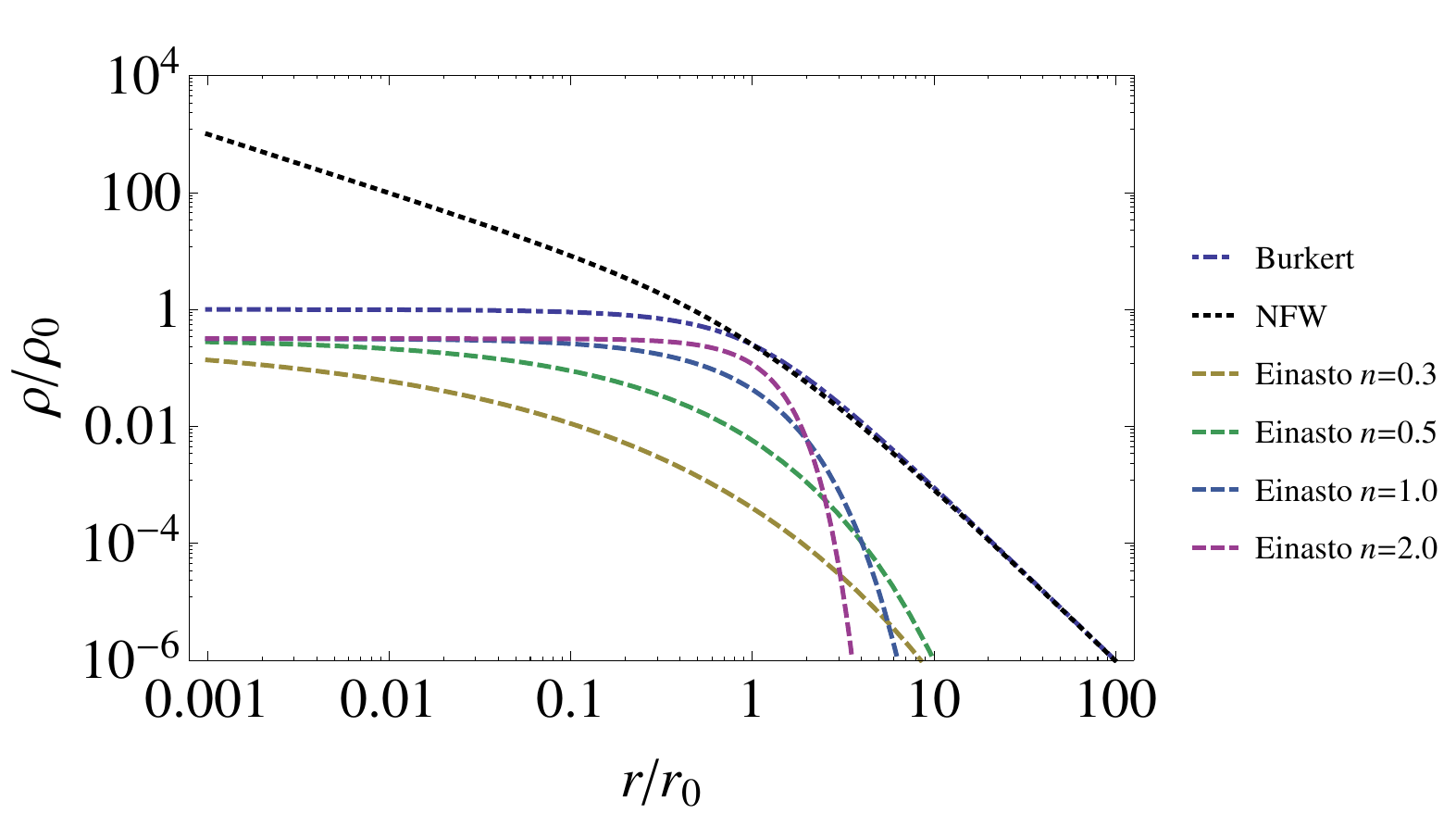}}
\vspace*{8pt}
\caption{Comparison of the density profiles for different phenomenological models
of dark matter distribution.}\label{fig1}
\end{figure}

After that, using the highest quality rotation curves available to date
obtained from The HI Nearby Galaxy Survey (THINGS) \citep{2008AJ....136.2563W,
2008AJ....136.2648D}, the Einasto dark matter halo model has been proposed as
the standard model for dark matter halos by \citet{2011AJ....142..109C}, as it
provides both cored and cusped distributions for different values of model
parameters (see Fig.~\ref{fig1}). In that work, the fundamental core-cusp
discrepancy is analyzed in detail for the whole sample of galaxies under study.
It is clearly shown that for the majority of the galaxies considered in the
sample, the cored halos (compatible with near unity Einasto indexes) are
preferred over the cuspy ones (these instead compatible with higher Einasto
indexes).

We present here a novel approach focusing on galactic structures and an
underlying microphysical component of Dark Matter. The model is built upon the
following general considerations: $1)$ the Dark Matter component is assumed to
be chargeless spin-1/2 fermions; $2)$ the configurations are described by
General Relativity; $3)$ the particles are assumed to be isothermal in
thermodynamic equilibrium (i.e. without the need of pre-fixing any cosmological
history). The theoretical fundament of this new approach is detailed in the
model of semi-degenerate self-gravitating fermions first introduced by
\citet{1990A&A...235....1G}, and more recently with applications to galactic
dark matter by \citet{2013PRL..R, 2013JKPS..A}. Our model is based in the
following main assumptions:
\begin{enumerate}
\item the problem of galactic cores and halos have to be addressed
    unitarily;
\item for definiteness we study the simplest problem of "bare" massive
    particles, neglecting at this stage all other interactions than the
    gravitational one and fulfilling only the Fermi--Dirac statistical
    distribution
\begin{equation}
    f=\frac{1}{\exp\left({\frac{\epsilon-\mu}{kT}}\right)+1}=
    \frac{1}{\exp{\left(\frac{\epsilon}{\beta mc^2}-\theta \right)}+1},
\end{equation}
where $\epsilon$ is kinetic energy of the particles, $\mu$ is chemical
potential, $T$ is the temperature, $k$ is Boltzmann constant and $c$ is the
speed of light. The mass of the particle ($m$), the temperature parameter
($\beta=kT/mc^2$) and the degeneracy parameter ($\theta=\mu/kT$) at the
center are the three free parameters of the model;
\item we consider zero total angular momentum and also neglect any effect
    of baryonic matter on the DM in the mathematical formulation.
\end{enumerate}

It is shown that in any such system the density at large radii scales as
$r^{-2}$ independently of the values of the central density, providing the flat
rotation curve \citep{1990A&A...235....1G, 2013PRL..R, 2013JKPS..A}.

The dark matter halos obtained in the new dark matter approach proposed here
share a common or universal feature which shed more light on the core-cusp
discrepancy, while providing a new mass scale to the dark matter candidate. Our
density profiles always favor a cored behaviour (without any cusp) in the
observed inner halo regions, given the quantum nature of the fermionic
particles. Another fundamental outcome of our model is the range of the DM
particle mass, which must be $m\gtrsim5$ keV in order to be in agreement with
typical halo sizes of the observed dwarf galaxies \citep{2013PRL..R}.

The theoretical formulation of \citet{2013PRL..R, 2013JKPS..A} is based on the
first principles physics and provides a physical complement to Einasto
phenomenological models. It also offers the necessity to approach the Dark
Matter distribution in galactic haloes with fermions with masses larger than
the above mentioned bound.

\begin{figure}[!h]
\centerline{\includegraphics[width=0.85\columnwidth]{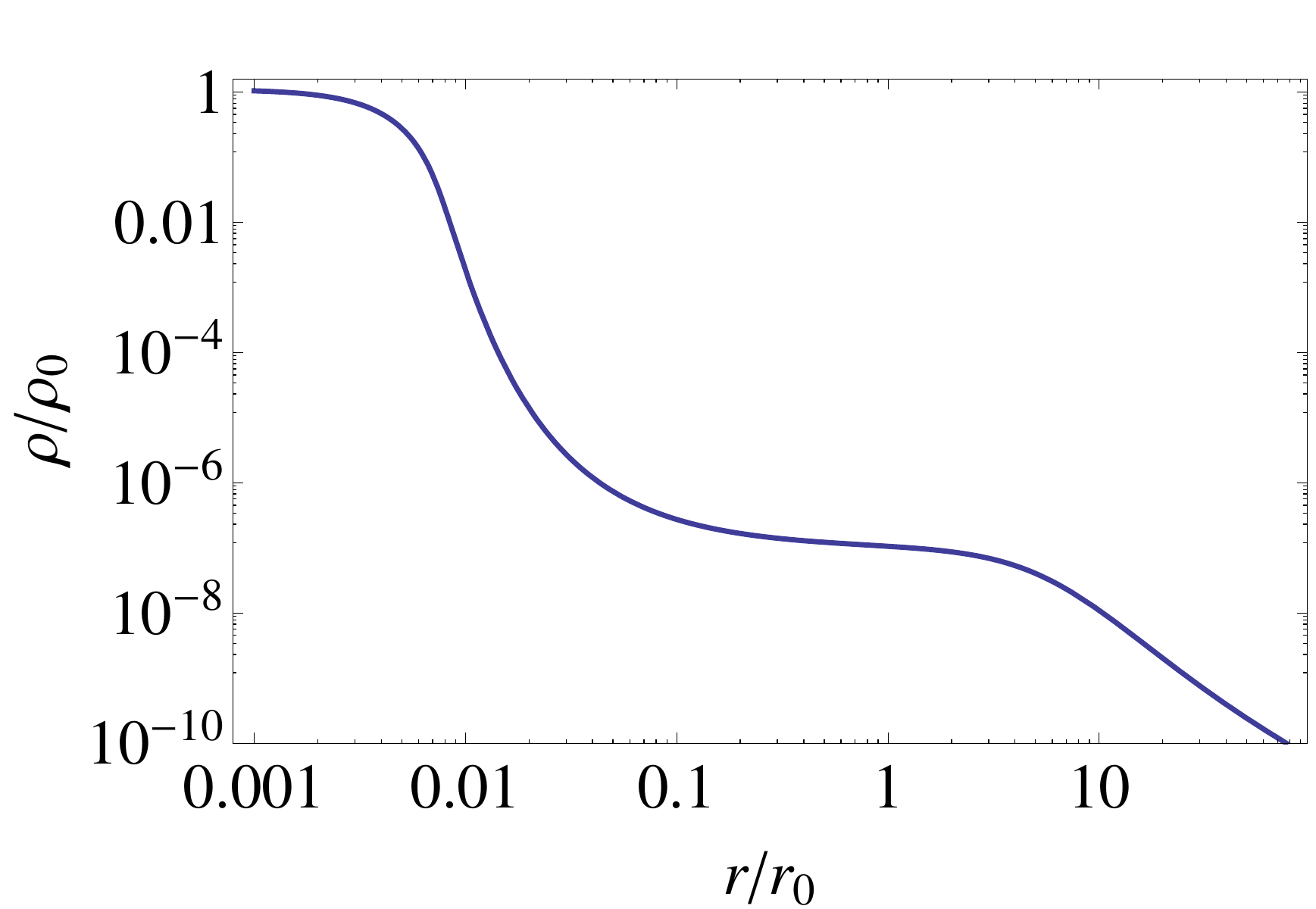}}
\vspace*{8pt}
\caption{Semidegenerate density profile in dimensionless units
for central degeneracy parameter $\theta_0=15$ and central temperature
parameter $\beta_0=10^{-10}$.}\label{fig12}
\end{figure}

The paper is structured as follows. We model the distribution of Dark Matter as
semidegenerate fully relaxed thermal self-gravitating general relativistic
fermionic solutions of \citet{1990A&A...235....1G}, see Sec.~\ref{model}. The
resulting density profiles provide flat rotation curve at large distances,
cored distribution of dark matter in the halo, and a massive degenerate core at
the very center, see Sec.~\ref{solutions}. We describe in Sec.~\ref{procedure}
the actual procedure of fitting of rotation curves, and then discuss the
results in Sec.~\ref{results}. Conclusions follow.

\section{Model equations}\label{model}

We consider self-gravitating system of fermions in thermal equilibrium
following \citet{1990A&A...235....1G} with occupation numbers given by
  \begin{equation}
    f(\epsilon)=\frac1{e^{\frac{\epsilon-\mu}{kT}}+1}.
  \end{equation}

Then equation of state reads
  \begin{gather}
    \rho=m\frac{g}{h^3}\int\frac{1+\epsilon/mc^2}%
    {e^{\frac{\epsilon-\mu}{kT}}+1}\,d^3p,\label{rho}\\
    P=\frac23 \frac{g}{h^3}\int
    \frac{(1+\epsilon/mc^2)^{-1}(1+\epsilon/2mc^2)\epsilon}%
    {e^{\frac{\epsilon-\mu}{kT}}+1}\,d^3p,\label{p}
  \end{gather}
where $g=2s+1$, $s$ is the spin of the particle, and integration is extended
over all 3-momentum space.

The Einstein equations for the spherically symmetric metric
\begin{equation}
g_{\mu \nu}={\rm diag}(e^{\nu},-e^{\lambda},-r^2,-r^2\sin^2\theta),
\end{equation}
where $\nu$ and $\lambda$ depend only on the radial coordinate $r$, together
with the thermodynamic equilibrium conditions of \citet{1930PhRv...35..904T}
and \citet{klein49}
\begin{equation}
e^{\nu/2} T=const\, , \qquad e^{\nu/2}(\mu+m c^2)=const,
\end{equation}
can be written in the dimensionless form of \citet{1990A&A...235....1G}
\begin{align}
	\frac{d\hat M}{d\hat r}&=4\pi\hat r^2\hat\rho, \label{OV1}\\
	\frac{d\theta}{d\hat r}&=-\frac{1-\beta_0(\theta-\theta_0)}{\beta_0}
    \frac{\hat M+4\pi\hat P\hat r^3}{\hat r^2(1-2\hat M/\hat r)},\label{eq:eqs2}\\
    \frac{d\nu}{d\hat r}&=\frac{\hat M+4\pi\hat P\hat r^3}{\hat r^2(1-2\hat M/\hat r)}, \\
    \beta_0&=\beta(r) e^{\frac{\nu(r)-\nu_0}{2}},\label{eq:tolman2}\\
    e^{-\lambda}&=1-\frac{2\hat M(\hat r)}{\hat r}\, .\label{OV2}
\end{align}
The following dimensionless quantities were introduced:
\begin{align}
&\hat r=r/\chi,\label{eq:dimensionlessr}\\
&\hat M=G M/(c^2\chi),\label{eq:dimensionlessM}\\
&\hat\rho=G \chi^2\rho/c^2,\\
&\hat P=G \chi^2 P/c^4,\label{eq:dimensionlessP}
\end{align}
where $\chi=2\pi^{3/2}(\hbar/mc)(m_p/m)$ is the characteristic length that
scales as $m^{-2}$, with $m_p=\sqrt{\hbar c/G}$ being the Planck mass, and the
temperature and degeneracy parameters, $\beta=k T/(m c^2)$ and $\theta=\mu/(k
T)$, respectively. The constants of the equilibrium conditions of Tolman and
Klein have been evaluated at the center $r=0$, which we indicate with a
subscript `0'.

The system of coupled differential equations (\ref{OV1}--\ref{OV2}) is solved
for initial conditions $M(0)=\nu(0)=0$ and given set of free parameters
$\beta_0$ and $\theta_0$, $m$ for each galaxy under study as detailed below.

\section{Properties of semidegenerate configurations}\label{solutions}

Galactic halos have to be necessarily composed from cold particles, so that
astrophysically relevant solutions will have temperature parameters
$\beta\ll1$. In this case, the general solution for semidegenerate
configurations ($\theta_0\gtrsim10$) present three different regions: an inner
degenerate compact core, an extended low-degenerate inner halo of almost
constant density and a non-degenerate outer halo with characteristic slope
$\rho\propto r^{-2}$ (see \cite{1990A&A...235....1G} and Fig.~\ref{fig12} for
additional details). The infinite mass of the configuration extended up to
spatial infinity is not a problem, because in reality it is limited by tidal
interactions with other galaxies, which introduce an energy cutoff into the
distribution function, see e.g. \cite{1992A&A...258..223I}. However this is not
important for the inner parts of the configuration we are interested in.

\begin{figure}
\center\includegraphics[width=0.85\columnwidth]{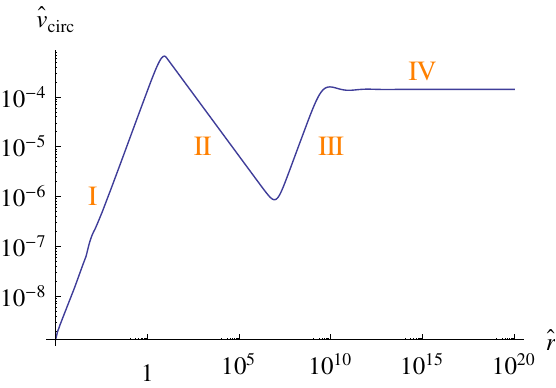}
\caption{\label{v}Dependence of $\hat v_{\text{circ}}=v_{\text{circ}}/c$
on dimensionless radius $\hat r$, $\beta=10^{-8},\ \theta_0=50$.}
\end{figure}

In order to understand the crucial properties of this equilibrium
configurations we plot the circular velocity of a test body in the metric
fulfilling Eqs.~(\ref{OV1}--\ref{OV2}) on Fig.~\ref{v}. There are indeed four
regions of the solution for circular velocity, each with characteristic slope.
The inner region I correspond to the degenerate core of almost constant
density, so that $v_{\text{circ}}\propto r$. For increasing values of radial
coordinate the inner halo follows, itself composed of two different regions. In
the region II the density of dark matter sharply decreases and this Keplerian
region is dominated by the mass of the degenerate core, and as a result
$v_{\text{circ}}\propto r^{-1/2}$. For yet increasing values of the radial
coordinate the density of dark matter reach an almost constant value giving
rise to a plateau, see Fig.~\ref{fig12}. As soon as the mass of the plateau
prevails over the mass of the core we have the region III where
$v_{\text{circ}}\propto r$. Finally in the region IV, after some oscillations
the circular velocity tends to a constant independent on $r$, corresponding to
a pure Boltzmannian regime and characteristic for the flat rotation curve of
outer halo.


We define the physical characteristics of each configuration as follows:
\begin{itemize}
\item The characteristic radius of the core $\hat r_c$ is given by $\hat
    v_{\text{circ}}(\hat r_c)=\mathrm{max}$ in region I.
\item $\hat M_{c}$ is the mass of the core given by $\hat M_c=\hat
    v^2_{\text{circ}}\hat r$ in the region II.
\item The characteristic radius of the inner halo $\hat r_h$ correspond to
    $\hat v_{\text{circ}}(\hat r_h)=\mathrm{max}$ in region III.
\item The characteristic mass of the inner halo $\hat M_h$ is given by
    $\hat M_h=\hat M(\hat r_h)$ just between the regions III and IV.
\end{itemize}

For the parameters in the region of $\theta_0\in[0,200],
\log\beta_0\in[-10,-5]$ we calculate a grid of models and extracted numerically
the physical characteristics mentioned above. Then we fit the obtained values
by different double parametric functions and find out the best fitting formulae
with the correspondent ($\beta_0$,$\theta_0$) dependence for the range of
$\theta_0\in[20,200]$. An interesting fact is that in our region of parameters
circular velocity $v_{\text{circ}}$ in the flat part of region IV (i.e.
$v_\infty$) is defined by temperature $\beta_0$ only. In the range of
astrophysically relevant parameters $\theta_0\in[10,200],
\log\beta_0\in[-10,-5]$ the scaling relation between circular velocity and
$\beta_0$ corresponds to the Boltzmannian relation between $v_{\text{circ}}$
and one-dimensional dispersion velocity $\sigma=\sqrt{kT/m}$ (see e.g.
\cite{1987gady.book.....B})
\begin{equation}\label{vinf}
\frac{v_\infty}{\text{km/s}}=\sqrt{2}c\sqrt{\beta_0}.
\end{equation}
We have here neglected general relativistic corrections which are very small in
these ranges of parameters, i.e. $e^{\nu(\infty)-\nu(0)}\approx1$, and then
$\beta(\infty)\approx\beta_0$ by equation (\ref{eq:tolman2}).

For the temperature and degeneracy free parameters in the range
$\log\beta_0\in[-10,-5]$, $\theta_0\in[20,200]$, respectively, we obtain the
following dimensionless scaling laws for core radius and mass\footnote{For
radii $r<r_c$ the configurations corresponds to region I, where $\rho(r)\approx
const$ and then from (\ref{eq:rcSL}) and (\ref{eq:McSL}), taking into account
(\ref{eq:dimensionlessr}) and (\ref{eq:dimensionlessM}), we can write
$\rho_c\propto M_c/r_c^3$ in terms of the chemical potential and particle mass
($\beta_0\theta_0=\mu_0/mc^2$) as $\rho_c\propto \mu_0^{3/2}m^{5/2}$. This
dependence is precisely the one of a fully degenerate non-relativistic Fermi
gas in presence of an external gravitational field, which is further coinciding
with a polytrope of index $n=3/2$ (see e.g. \citet{1983bhwd.book.....S}).}
\begin{gather}
\hat r_{c}=0.226(\beta_0\theta_0)^{-1/4}, \label{eq:rcSL}\\
\hat M_{c}=0.234(\beta_0\theta_0)^{3/4}. \label{eq:McSL}
\end{gather}
However, the core region is typically very small and is not constrained by
empirical data of the THINGS sample considered here. Moreover, the mass
contribution of regions I and II to the total mass $M_h$ at the end of region
III is $\lesssim10^{-2}$ in our parameter range as shown in \cite{2013PRL..R},
indicating that only regions III and IV are the relevant ones to be used in the
fitting procedure against the data.

At this point it is important to emphasize that the theoretical treatment used
here to fit dark matter halos applies for any core size, even for the ones
which are close to its critical mass, $M_{cr}\sim 10^9 M_\odot$, as studied in
\cite{2014JKPS..A}, where a relativistic treatment is mandatory. Even though
the regions I-II-III-IV corresponding to the sample considered here can be well
explained in terms of non-relativistic physics, the general relativistic
approach has been used for formal correctness, only giving negligible
corrections.

Dimensionless halo radius and mass have different scalings, they are
proportional not to $\theta_0^\alpha$, but to $\alpha^{\theta_0}$
\begin{gather}
\hat r_h=0.953\beta_0^{1/4}(1.445)^{\theta_0}, \label{eq:rhalo}\\
\hat M_h=2.454\beta_0^{3/4}(1.445)^{\theta_0}. \label{eq:Mhalo}
\end{gather}

Formulas (\ref{eq:rhalo}--\ref{eq:Mhalo}) represent perfect scalings in the
region of parameters considered above, which involves also the scaling of the
whole rotational curves in regions III and IV; formula (\ref{vinf}) shows a
perfect scaling in the flat part of the rotation curve for region IV. Moreover,
the Newtonian expression for the dimensionless circular velocity $\hat
v_{circ}^2(\hat r)=\hat M/\hat r$ is perfectly suitable in the physical region
under consideration. The expression for maximal rotation velocity in the halo
is thus obtained from (\ref{eq:rhalo}) and (\ref{eq:Mhalo}) and reads
\begin{equation}
\hat v^2_h(\hat r_h(\beta_0,\theta_0))=2.575\beta_0^{1/2}.
\label{eq:vhalo}
\end{equation}

In next section we explain the fitting procedure with the use of the halo
scaling laws for regions III and IV obtained here.

\section{Observed rotation curves and fitting procedure}\label{procedure}

In 2008, a sample of 34 nearby (closer than 15 Mpc) spiral and irregular
galaxies (Sb to Im) were observed with The HI Nearby Galaxy Survey (THINGS)
\citep{2008AJ....136.2563W}. These observations allowed to obtain the highest
quality rotation curves available to date due to the high spatial and velocity
resolution of THINGS. Then a sub-sample of these rotation curves, corresponding
to 19 rotationally dominated and undisturbed galaxies, were combined with
information on the distribution of gas and stars by \citet{2008AJ....136.2648D}
to construct mass models for the dark matter component of the sample. These
models finally were used to quantify the dark matter contribution for each
galaxy by using the following formula
\begin{equation}
V^2_{obs}=V_{gas}^2+\Upsilon_*V_*^2+V_{DM}^2,
\label{eq:massmodel}
\end{equation}
which relates the observed input curves of $V_{obs}$, $V_{gas}$ and $V_{*}$,
defined below, with the dark matter rotation curve $V_{DM}$ to be determined
from the known input data once the the mass-to-light ratio $\Upsilon_*$ is
provided.

The total and gas observed rotation curves $V_{obs}$ and $V_{gas}$,
respectively, were both obtained from the THINGS data: the first was obtained
from velocity fields analysis and the second from the neutral hydrogen (HI)
distribution maps, as described in \citet{2008AJ....136.2648D}. Instead, each
stellar (light) rotation curve $V_{*}$ is obtained from the corresponding
stellar distribution observed in the K band (i.e. at $3.6\ \mu$m) by the
\textit{Spitzer} Infrared Nearby Galaxy Survey (SINGS), independent of THINGS,
and described in \citet{2008AJ....136.2648D} and references therein. Finally,
the mass-to-light ratio $\Upsilon_*^K$ was used to determine the rotation curve
associated with the stellar mass distribution from that of the measured light.

At this point it is relevant to further emphasize the underlying hypothesis to
which equation (\ref{eq:massmodel}) is subject to. This is, each baryonic
rotation velocity $V_{gas}$ and $V_{*}$ was calculated from the correspondent
observed baryonic mass density distribution, and was defined as the velocity
that each component would induce on a test particle in the galactic plane as if
they were isolated of any external influence.

In \citet{2008AJ....136.2648D} equation (\ref{eq:massmodel}) was applied to
test the cuspy Navarro-Frenk-White and cored pseudo-ISO dark matter models
against data as follows: the (squared) rotation curves of the baryonic
components (after appropriate scaling with $\Upsilon_*^K$) were subtracted from
the (squared) observed rotation curve $V^2_{obs}$ to apply a reduced $\chi^2$
fitting procedure in order to find the best fitting free parameters for each
dark matter model. Soon after, the same analysis was extended further to
Einasto dark matter profiles by \citet{2011AJ....142..109C}, concluding that
the Einasto model provides the best match to the observed rotation curves when
compared with NFW and pseudo-ISO models with empirical fixed values for
$\Upsilon_*^K$ for two different stellar initial mass functions (IMFs).

Here we propose a different dark matter halo model, which is neither based on
numerical N-body simulations nor on phenomenological model proposals, but
relies on the underlying microphysical composition of the dark matter
candidate, as explained in former sections.

Thus, analogously to \citet{2008AJ....136.2648D} and
\citet{2011AJ....142..109C} we use the HI high resolution observations of
galaxies from THINGS survey \citep{2008AJ....136.2563W}. We analyze here the
sample of 16 rotationally dominated and undisturbed galaxies presented both in
\citet{2008AJ....136.2648D} and \citet{2011AJ....142..109C}, listed below in
Table 1.

Regarding the rotation curves data of the baryonic components, we consider the
contributions of the gas, the stellar disk and a spherical stellar bulge $V_b$
as given in \citet{2011AJ....142..109C}. The halo rotation velocity
corresponding to the spherical dark matter component is taken from the two
parametric scaling, see (\ref{eq:rhalo}--\ref{eq:Mhalo}), which we name from
now on as the fermionic dark matter velocity profile $V_{f}(r)$.

Once each component is provided, we make use of the equation analogous to
(\ref{eq:massmodel})
\begin{equation}
V^2_{obs}=V_{gas}^2+\Upsilon_*V_*^2+V_{b}^2+V_{f}^2,
\label{eq:massmodel2}
\end{equation}

With all the baryonic velocity terms ($V_{gas}^2$, $\Upsilon_*V_*^2$ and
$V_b^2$) as observational inputs, we fit the HI observed rotation curve
$V^2_{obs}$ by Levenberg--Marquardt nonlinear least-squares algorithm, in
complete analogy as done in \citet{2011AJ....142..109C}.

We did not take into account the contribution of molecular gas because the
total gas surface densities are dominated by atomic gas for the majority of the
sample, as explained in \citet{2011AJ....142..109C} and references therein.
Total rotation curve was taken from \citet{2008AJ....136.2648D}. We have not
considered models with free mass-to-light ratios, deferring it to a future
paper (in preparation). Instead following \citet{2011AJ....142..109C} we have
adopted the fixed mass-to-light ratios of stellar populations with a bursty
star formation history with a Kroupa IMF. We choose this IMF instead of the
diet-Salpeter IMF, also considered in \citet{2011AJ....142..109C} and
\citet{2008AJ....136.2648D}, as it generally provides better agreement with
observations for rotation curves (see Fig. 5 of \cite{2011AJ....142..109C}),
and in some cases the Salpeter IMF leads to rotational velocities due to
stellar component only already in pronounced excess over observed total
rotational velocity (see, for example, cases of
NGC3521 and NGC5055 at Fig. 3 of \cite{2011AJ....142..109C}). 

Together with the dark matter profile of semidegenerate configurations with
particle mass $m=10$~keV/$c^2$ and varying $\theta_0$ and $\beta_0$, and for
the sake of comparison, the following profiles were also used for fitting:
\begin{itemize}
\item Cored profiles with central density $\rho_0$ and characteristic
    radius $r_0$:
\begin{itemize}
\item pseudo-isothermal sphere profile
\begin{equation}\label{pISO}
    \rho_{DM}(r)=\rho_{0}\frac{r_0^2}{r^2+r_{0}^2},
\end{equation}
\item Burkert profile
\begin{equation}\label{Burkert}
    \rho_{DM}(r)=\rho_{0}\frac{r_{0}^3}{(r_0+r)(r_0^2+r^2)}.
\end{equation}
\end{itemize}
\item Cusped profiles with characteristic radius $r_{-2}$ where the density
    profile has a (logarithmic) slope of $-2$ (the ''isothermal'' value)
    and $\rho_{-2}$ as the local density at that radius. In the case of
    Einasto profiles a third parameter is needed, the Einasto index $n$
    which determines the shape of the profile.
\begin{itemize}\item Navarro--Frenk--White profile
\begin{equation}\label{NFW}
    \rho_{DM}(r)=4\rho_{-2}\frac{r_{-2}}{r}\left(\frac{r_{-2}}{r+r_{-2}}\right)^2,
\end{equation}
\item Einasto profile\footnote{The family of Einasto profiles with
    relatively large indices $n>4$ are identified with cuspy halos,
    while low index values $n<4$ presents a cored-like behaviour
    \citep{2011AJ....142..109C}. The lower the $n$ the more cored-like
    the halo profile.}
\begin{equation}\label{Einasto}
    \rho_{DM}(r)=\rho_{-2}\exp\left\{-2n\left[\left(\frac{r}{r_{-2}}\right)^{1/n}-1\right]\right\}.
\end{equation}
\end{itemize}
\end{itemize}

\section{Results and discussion}\label{results}

In this section, we compare the fits of rotation curves by the different models
considered in the last section. As we have to compare models with different
number of parameters, which are not nested into each other, we use Bayesian
Information Criterion (BIC) introduced by \citet{Schwarz1978}. It provides a
penalty to models with larger number of parameters to check what of them is
more likely to be correct. Model with minimum BIC value is preferred. For the
models with the same number of parameters, BIC is equivalent to $\chi^2$
criterion.

\begin{table*}{\tiny\noindent
\caption{Results of fitting}\label{fitres}
\begin{tabular}{| c |r@{${}\pm{}$}lr@{${}\pm{}$}lcc|
r@{${}\pm{}$}lr@{${}\pm{}$}lcc|}
\hline
& \multicolumn{6}{c|}{Semidegenerate}&
\multicolumn{6}{c|}{Burkert}\\
\textbf{Galaxy}&
\multicolumn{2}{c}{$\beta, 10^{-8}$}&
\multicolumn{2}{c}{$\theta_0^*$}&
$\chi^2_r$ & BIC&
\multicolumn{2}{c}{$r_0$, kpc}&
\multicolumn{2}{c}{$\rho_0,\ 10^{-3}\ M_{\odot}/\text{pc}^3$}&
$\chi^2_r$ & BIC\\\hline
 NGC2366& 0.99 & 0.02 & 24.22& 0.08 & 0.10 & \emph{27} &
          2.2  &  0.2 &  43  &  10  & 0.12 & 35 \\
 NGC2403& 7.10 & 0.06 & 27.43& 0.07 &  2.7 & 902&
          4.08 & 0.06 &  83  &  3   &  2.3 & 866\\
 NGC2841& 10.8 & 0.4  & 32.04& 0.17 &  2.6 & \emph{366}&
          20.6 &  0.9 &  5.2 &  0.5 &  2.7 & 370\\
 NGC2903& 16.33& 0.09 & 26.88& 0.06 & 0.66 & \emph{238}&
          2.89 & 0.06 &  388 &  18  &  1.1 & 283\\
 NGC2976& 9    & 3    & 28.8 & 0.4  & 0.44 & \emph{93} &
          20   &  20  & 40   & 150  & 0.49 & 97 \\
 NGC3031& 9.7  & 0.3  & 26.6 & 0.2  &  3.9 & 470&
          2.63 &  0.10&  270 &  20  &  3.9 & 468\\
 NGC3198& 7.44 & 0.07 & 28.66& 0.08 & 1.06 & 254&
          6.32 &  0.18&  36  &  2   & 0.99 & \emph{248}\\
 IC2574 & 2.00 & 0.06 & 27.96& 0.09 & 0.28 & 167&
          8.0  &  0.6 &  7.2 &  1.2 & 0.10 &  67\\
 NGC3521& 8.9  & 0.7  & 28.3 & 0.4  &  4.1 & \emph{437}&
          5.4  &  0.4 &  60  &  8   &  4.2 & 437\\
 NGC3621& 7.35 & 0.11 & 28.70& 0.08 &  3.0 & 496&
          6.48 &  0.12&  34.3&  1.3 &  2.5 & 475\\
 NGC4736& 3.01 & 0.13 & 22.4 & 0.5  & 2.1  & 302&
          0.84 &  0.07&  870 & 160  & 1.9  & 296\\
 DDO154 & 0.791& 0.017& 24.37& 0.10 & 0.84 & 172&
          2.32 &  0.10&  29  &  3   & 0.62 & 153\\
 NGC5055& 8.35 & 0.13 & 30.99& 0.12 &  1.9 & 708&
          14.3 &  0.5 &  8.0 &  0.5 &  2.0 & 719\\
 NGC6946& 8.9  & 0.3  & 26.9 & 0.2  & 16.4 & 993&
          3.48 &  0.08&  146 &  7   &  15.7& 985\\
 NGC7331& 11.7 & 0.2  & 30.10& 0.12 & 0.45 & 226&
          9.7  &  0.7 &  25  &  4   & 0.41 & 214\\
 NGC7793& 4.44 & 0.16 & 25.85& 0.13 & 3.7  & 287&
          2.60 &  0.07& 130  & 8    & 3.5  & 284\\\hline
\end{tabular}\\
\begin{tabular}{| c |r@{${}\pm{}$}lr@{${}\pm{}$}lcc|
r@{${}\pm{}$}lr@{${}\pm{}$}lcc|
r@{${}\pm{}$}lr@{${}\pm{}$}lr@{${}\pm{}$}lcc|}
\hline
& \multicolumn{6}{c|}{Navarro--Frenk--White}&
\multicolumn{6}{c|}{Pseudo-ISO}\\
\textbf{Galaxy}&
\multicolumn{2}{c}{$r_{-2}$, kpc}&
\multicolumn{2}{c}{$\rho_{-2},\ 10^{-3}\ M_{\odot}/\text{pc}^3$}&
$\chi^2_r$ & BIC&
\multicolumn{2}{c}{$r_0$, kpc}&
\multicolumn{2}{c}{$\rho_0,\ 10^{-3}\ M_{\odot}/\text{pc}^3$}&
$\chi^2_r$ & BIC\\\hline
 NGC2366& 200  & 1100 &  0.02&  0.24&  1.1 & 117&
          1.29 & 0.17 &  40  &  11  &  0.15&  42\\
 NGC2403& 11.2 &  0.3 &  2.86&  0.17&  0.7 & 575&
          1.56 &  0.04&  144 &  8   &  1.2 & 703\\
 NGC2841& 150  &  30  &  0.05&  0.02&  3.8 & 402&
          12.5 &  0.7 &  4.6 &  0.6 &  2.8 & 374\\
 NGC2903& 4.75 &  0.16&   34 &  2   &  1.8 & 334&
          0.53 &  0.05&  2300&  500 &  3.9 & 406\\
 NGC2976& 900  & 40000& 0.009&  1   &  2.1 & 158&
          9    &  18  &  30  & 180  &  0.49& 98 \\
 NGC3031& 4.9  &  0.4 &  20  &   3  &  4.0 & 472&
          0.82 &  0.1 &  690 &  170 &  4.3 & 480\\
 NGC3198& 16.5 &  0.9 & 1.37 &  0.15&  2.0 & 306&
          2.7  &  0.14&  51  &  5   &  1.2 & 266\\
 IC2574 & 500  &  1300& 0.007& 0.045&  1.5 & 331&
          5.1  &  0.4 &  6.3 &  1.1 & 0.11 &  70\\
 NGC3521& 18   &  3   &  1.5 &  0.4 &  5.1 & 457&
          2.4  &  0.3 &  78  &  19  &  4.2 & 439\\
 NGC3621& 123  &  17  &  0.08&  0.02&  5.9 & 579&
          2.81 &  0.09&  49  &  3   &  1.1 & 377\\
 NGC4736& 1.25 &  0.16&  90  &  20  & 1.9  & 296&
          0    &  0.07&  0   & ND   & 2.3  & 311\\
 DDO154 & 14   &  2   &  0.35& 0.11 & 1.03 & 184&
          1.22 &  0.07&  32  &  4   & 0.48 & 138\\
 NGC5055& 48   &  4   &  0.20&  0.03&  3.0 & 795&
          7.8  &  0.4 &  7.7 &  0.8 &  2.5 & 759\\
 NGC6946& 9.3  &  0.4 &  5.3 &  0.5 &  10.4& 915&
          0.66 &  0.03&  870 &  70  &  11.0& 923\\
 NGC7331& 3000 &40000 & 0.004& 0.087& 0.48 & 233&
          5.   &  0.5 &  28  &  6   & 0.32 & 190\\
 NGC7793& 17.0 & 1.9  & 1.4  & 0.3  & 4.1  & 294&
          1.47 & 0.05 & 126  & 10   & 4.0  & 293\\\hline
\end{tabular}
\begin{tabular}{| c |r@{${}\pm{}$}lr@{${}\pm{}$}lr@{${}\pm{}$}lcc|}
\hline
& \multicolumn{8}{c|}{Einasto}\\
\textbf{Galaxy}&
\multicolumn{2}{c}{$r_{-2}$, kpc}&
\multicolumn{2}{c}{$\rho_{-2},\ 10^{-3}\ M_{\odot}/\text{pc}^3$}&
\multicolumn{2}{c}{$n$}&
$\chi^2_r$ & BIC\\\hline
 NGC2366& 2.9  &  0.4 &  6.86&  0.04&  0.9 & 0.3& 0.13&  39 \\
 NGC2403& 13.6 &  1.3 &  1.98&  0.05&  4.9 & 0.4&  0.6&  \emph{570}\\
 NGC2841& 24.5 &  0.6 & 1.091& 0.006& 0.54 &0.08&  2.5&  367\\
 NGC2903& 5.33 &  0.15&28.983& 0.005&  2.9 & 0.2&  1.6&  326\\
 NGC2976& 70000&   ND & 0.019& ND   &  4.0 & 70 & 0.49&  100\\
 NGC3031& 4.81 &  0.09&30.159& 0.002& 0.56 &0.07&  3.2&  \emph{452}\\
 NGC3198& 11.5 &  0.4 & 3.029& 0.006& 1.80 &0.17&  1.1&  257\\
 IC2574 & 8.6  & 1.2  &  1.57&  0.05&  0.7 & 0.2&  0.09&  \emph{62}\\
 NGC3521& 9.3  &  0.5 & 6.27 &  0.02&  1.2 & 0.3&  4.3&  443\\
 NGC3621& 37   &  10  &  0.3 &  0.4 &  6.1 & 0.9&  0.57&  \emph{296}\\
 NGC4736& 1.73 &  0.16& 56.78&  0.03&  2.0 & 0.5&  1.8&  \emph{295}\\
 DDO154 & 4.9  &  0.7 & 1.95 &  0.03&  2.0 & 0.3&  0.31&  \emph{114}\\
 NGC5055& 21.4 &  0.4 & 1.464& 0.002&  0.39&0.04&  1.2&  \emph{626}\\
 NGC6946& 50   &  40  &  0.2 &  4   &  15  &  4 &  8.5&  \emph{883}\\
 NGC7331& 1000 &  8000&  0.002&  900&  11  & 21 & 0.23&  \emph{158}\\
 NGC7793& 3.65 & 0.14 &19.217& 0.007& 0.97 &0.09&  3.1&  \emph{279}\\\hline
\end{tabular}}\\\ \\
{\small ND means not constrained model parameters}
\end{table*}

\begin{figure*}[p]
  \begin{tabular}{ccc}
  \emph{NGC2366} & NGC2403 & \emph{NGC2841}\\
  \includegraphics[width=0.3\textwidth]{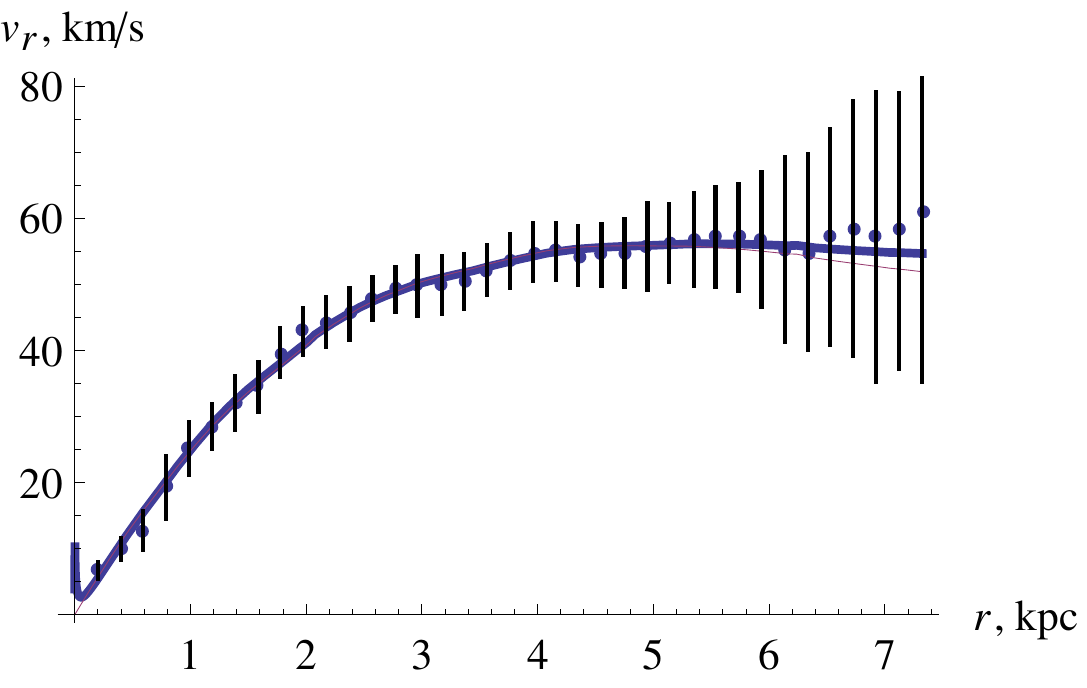} &
  \includegraphics[width=0.3\textwidth]{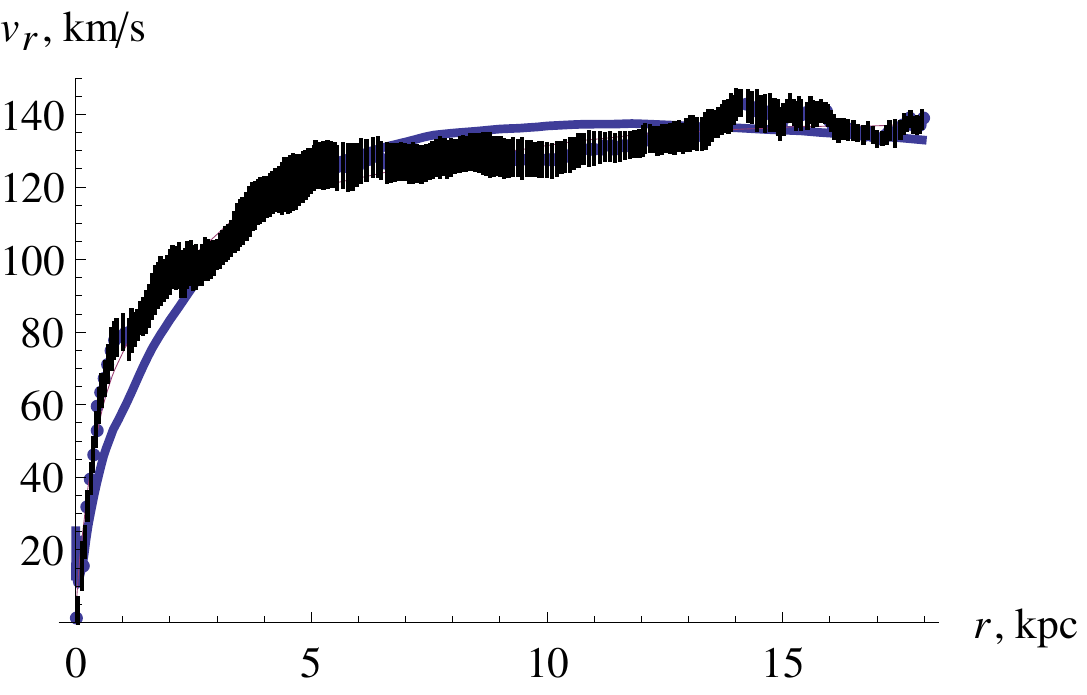} &
  \includegraphics[width=0.3\textwidth]{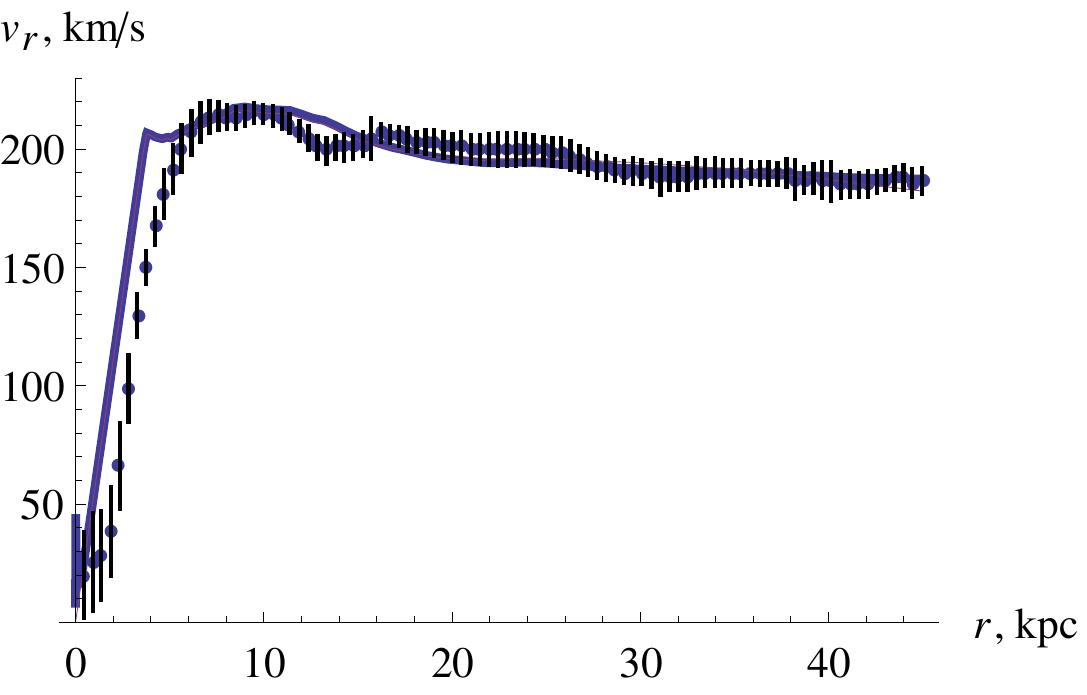} \\
  \emph{NGC2903} & \emph{NGC2976} & \textbf{NGC3031}\\
  \includegraphics[width=0.3\textwidth]{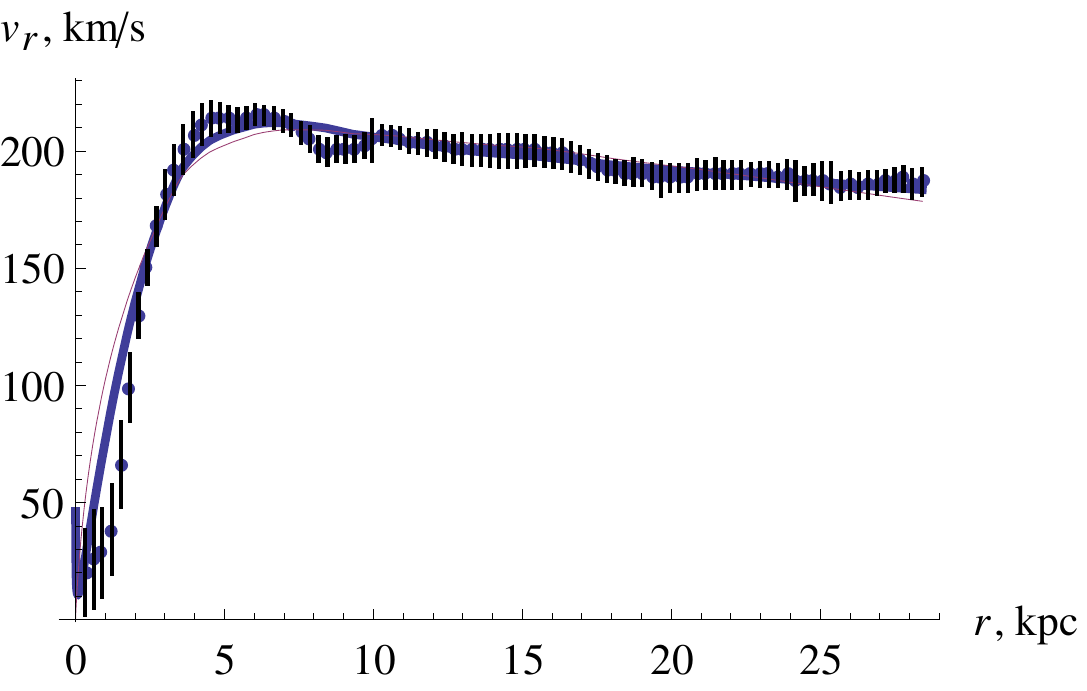} &
  \includegraphics[width=0.3\textwidth]{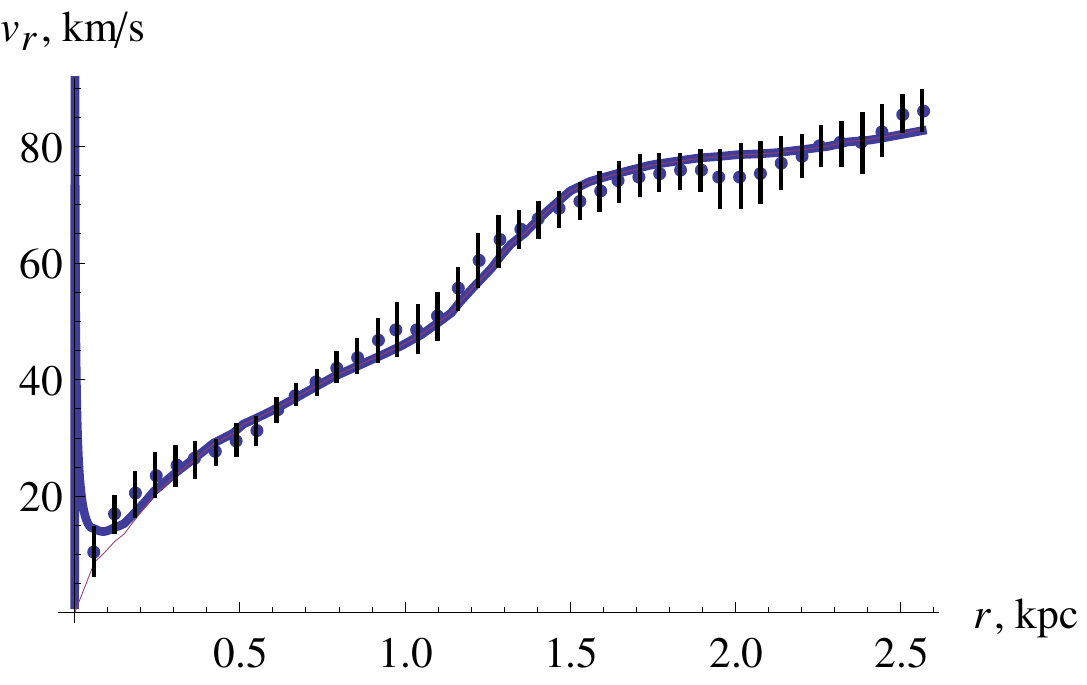} &
  \includegraphics[width=0.3\textwidth]{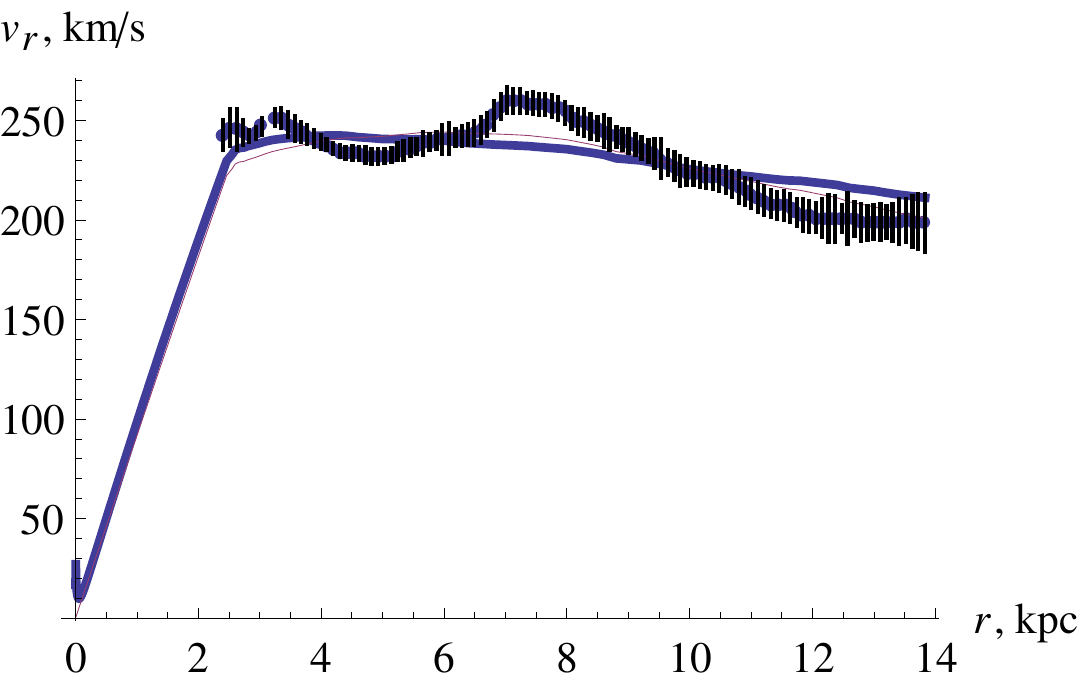} \\
  \emph{NGC3198} & IC2574 & \emph{NGC3521}\\
  \includegraphics[width=0.3\textwidth]{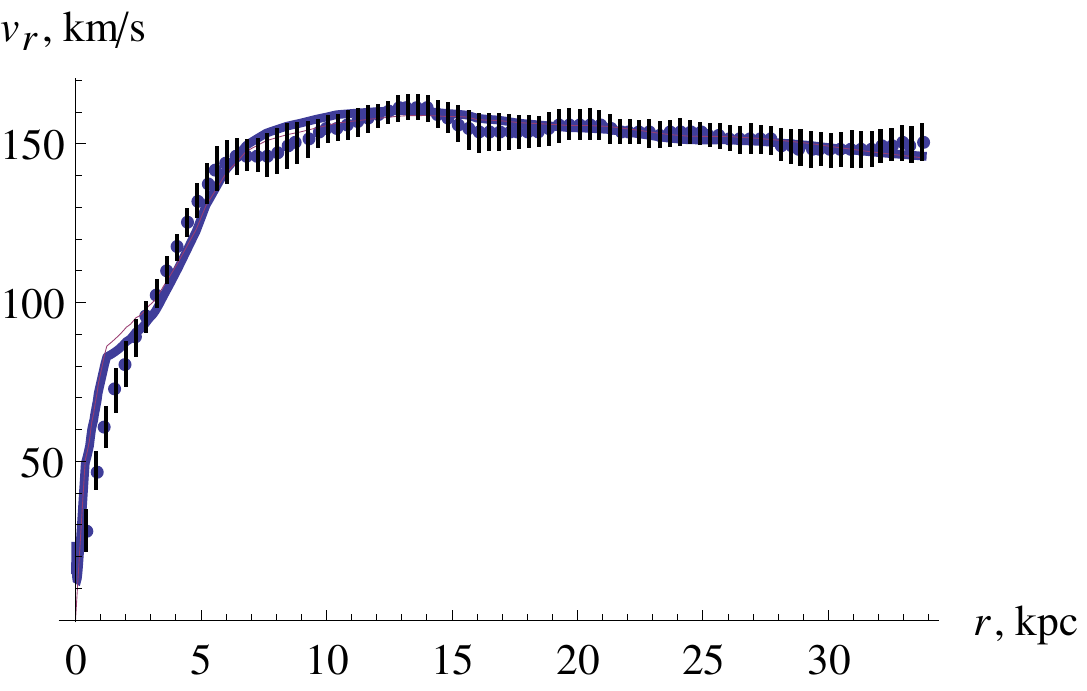} &
  \includegraphics[width=0.3\textwidth]{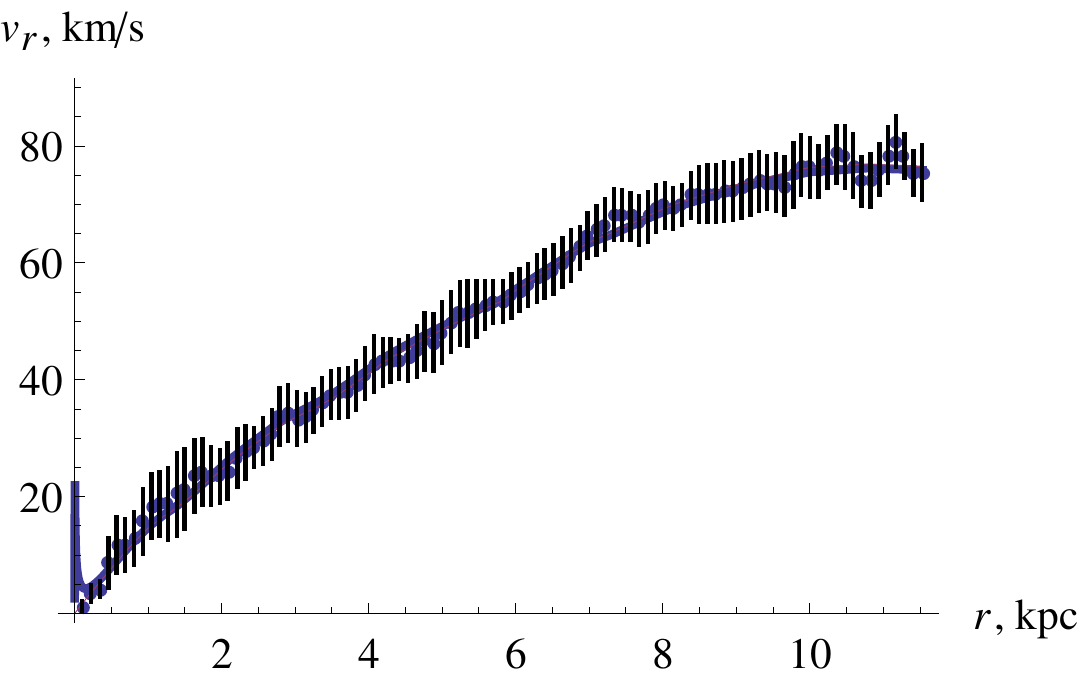} &
  \includegraphics[width=0.3\textwidth]{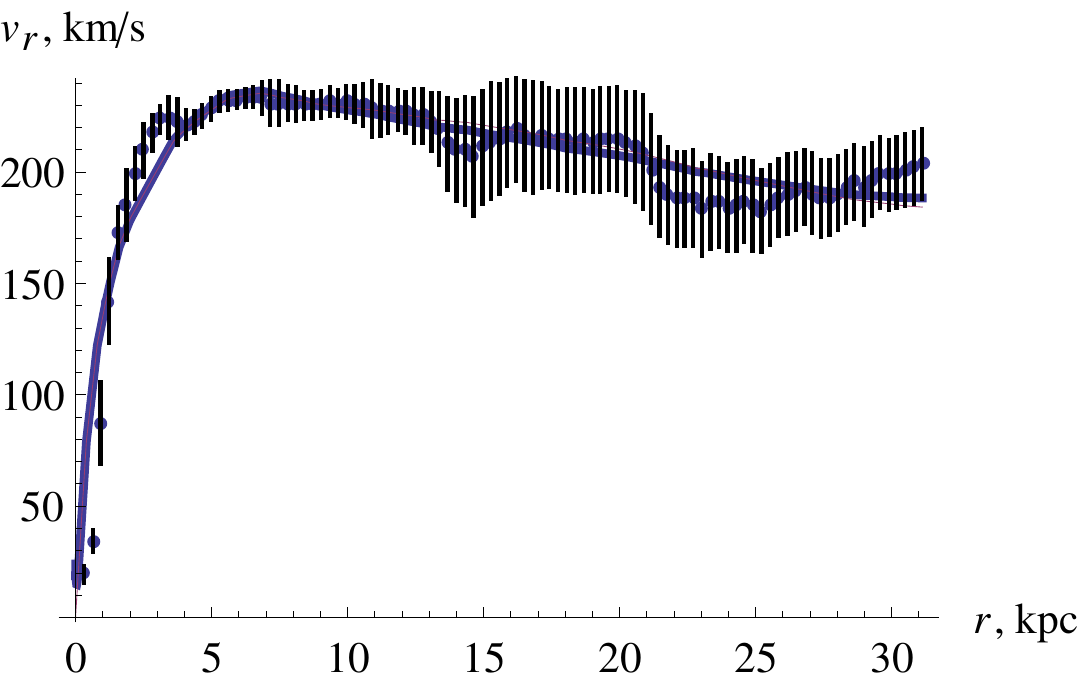} \\
  NGC3621 & \textbf{NGC4736} & DDO154\\
  \includegraphics[width=0.3\textwidth]{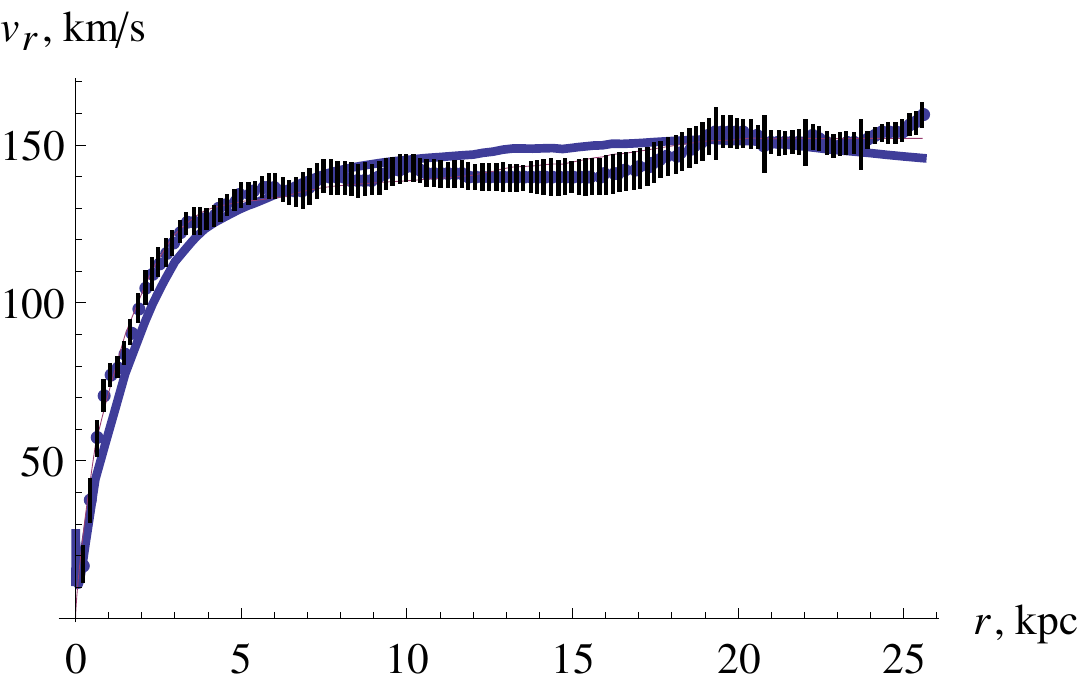} &
  \includegraphics[width=0.3\textwidth]{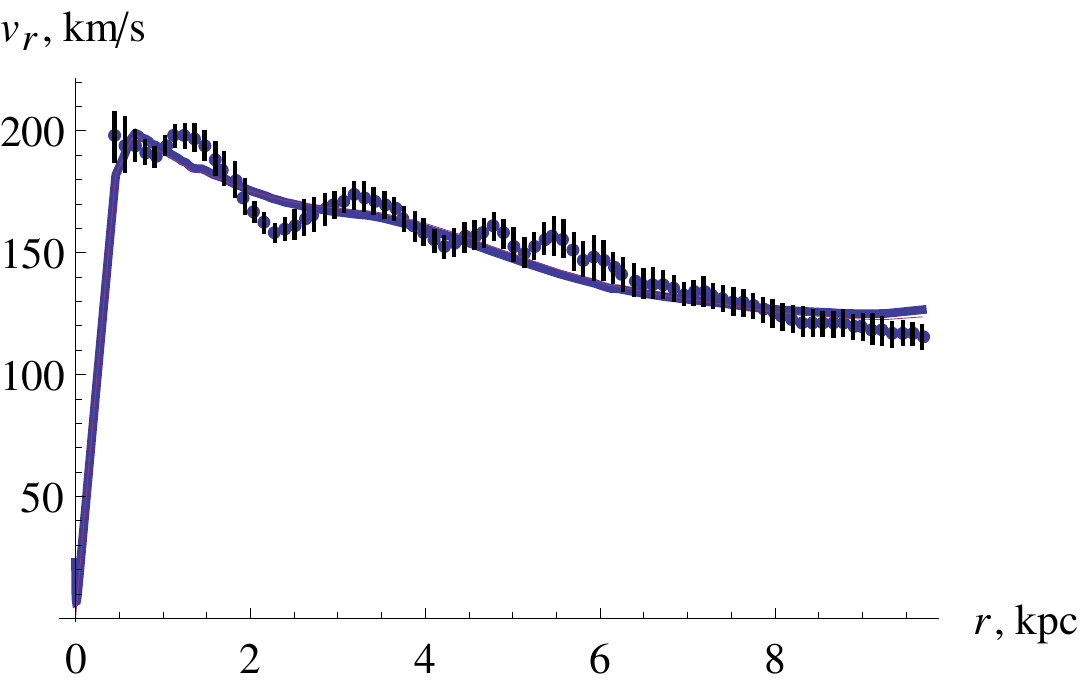} &
  \includegraphics[width=0.3\textwidth]{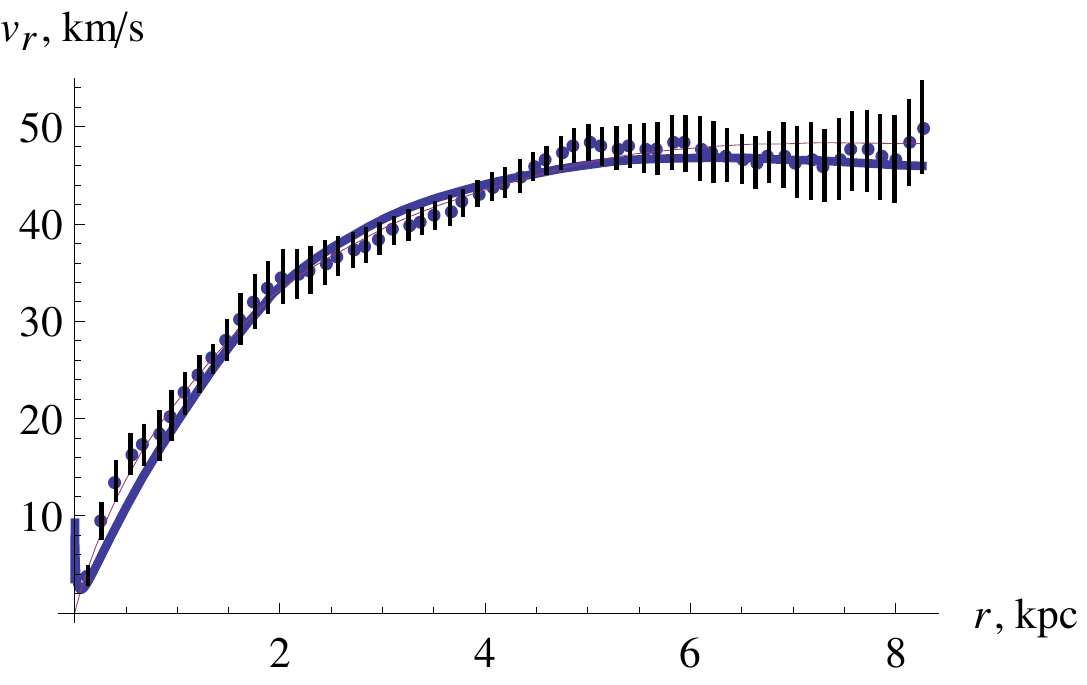} \\
  NGC5055 & NGC6946 & NGC7331\\
  \includegraphics[width=0.3\textwidth]{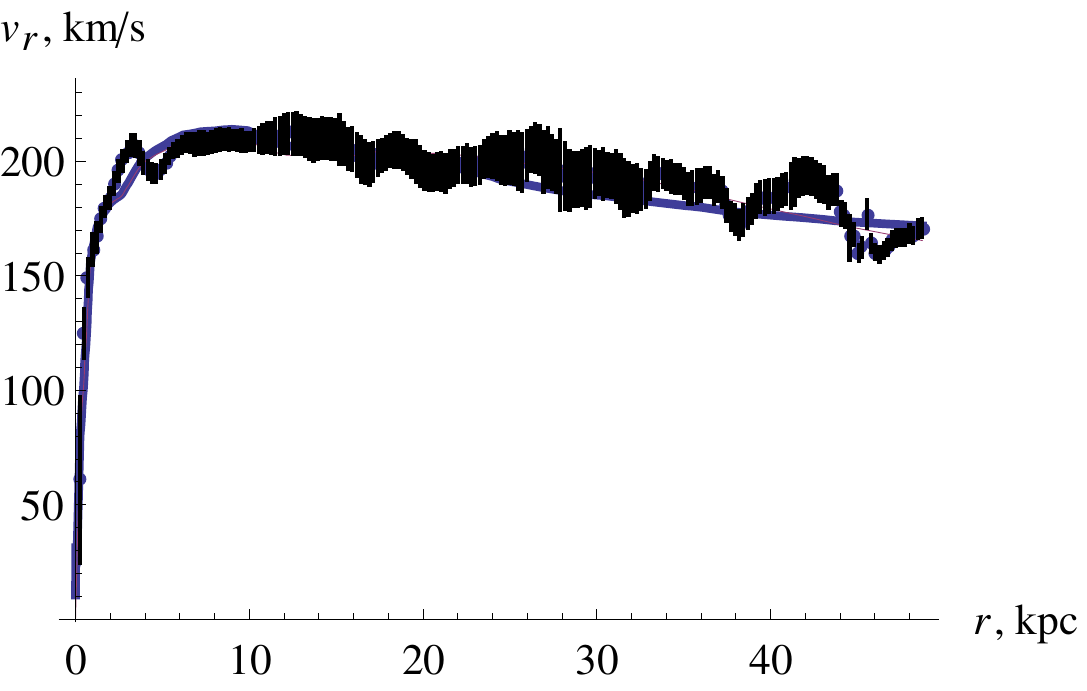} &
  \includegraphics[width=0.3\textwidth]{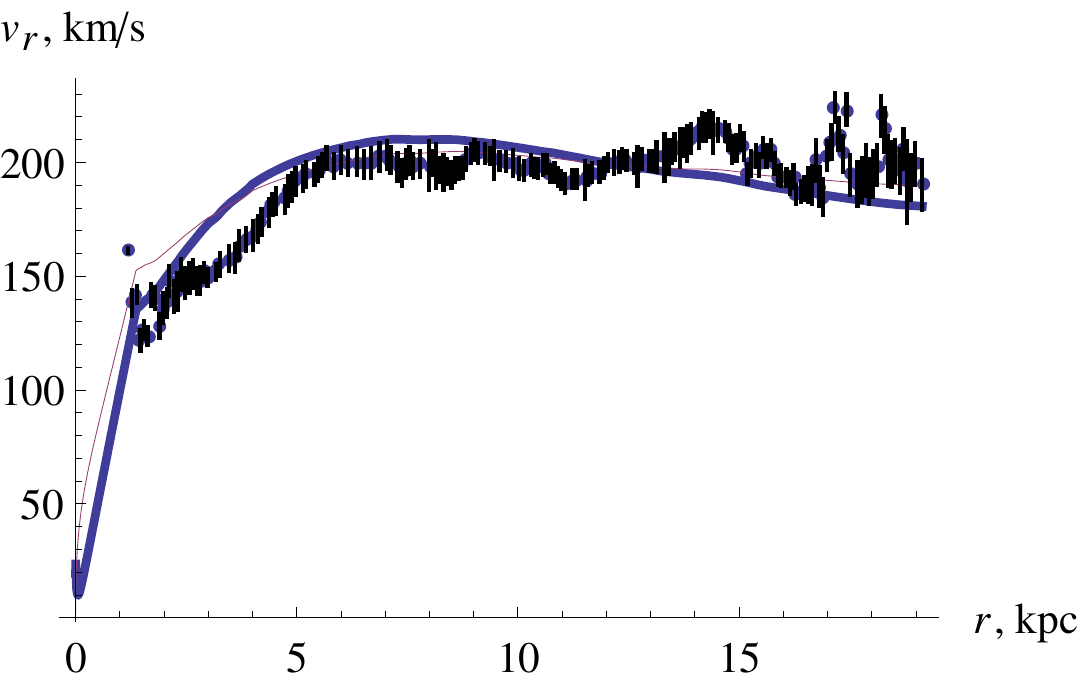} &
  \includegraphics[width=0.3\textwidth]{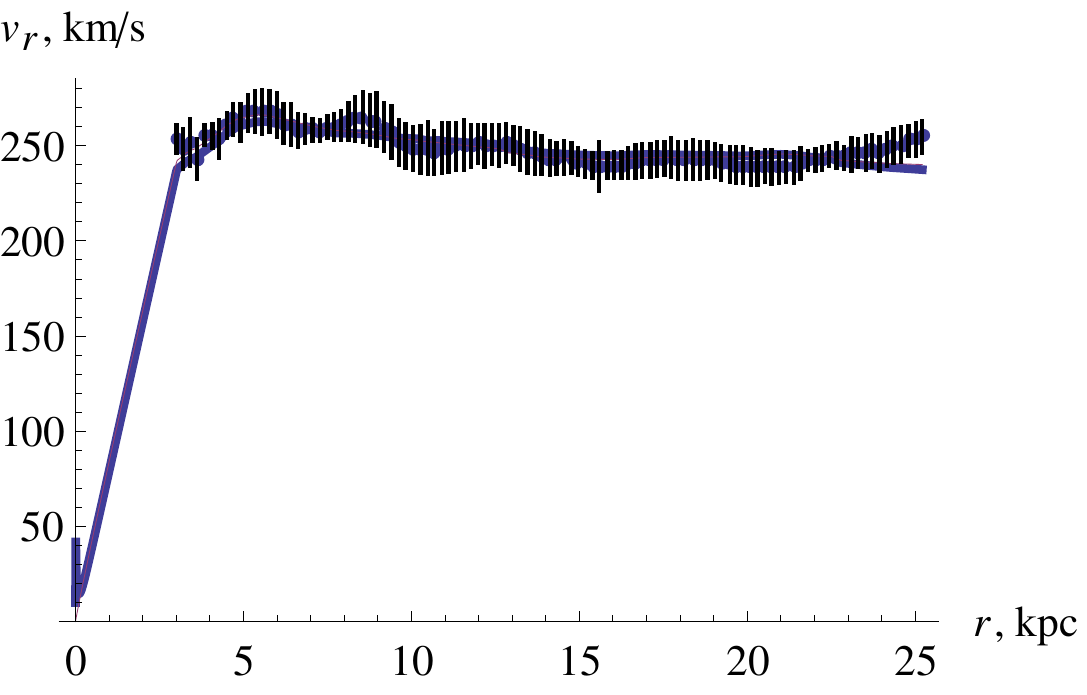} \\
  & \textbf{NGC7793} & \\
  & \includegraphics[width=0.3\textwidth]{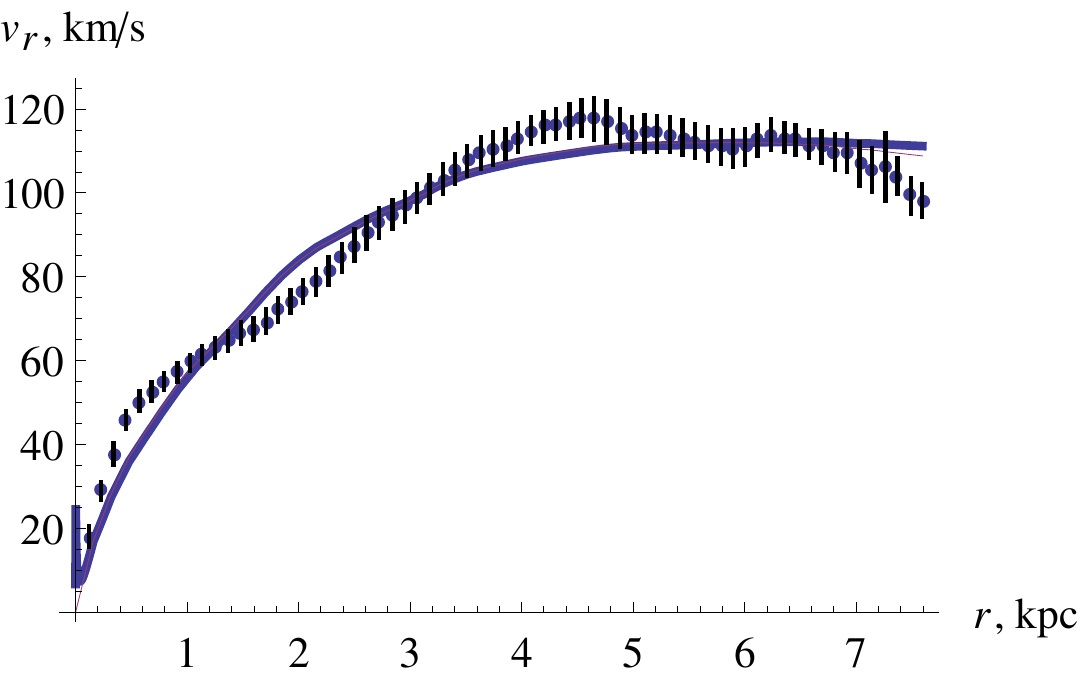} & \\
  \end{tabular}
  \caption{Rotational curves $v_{r}(r)$ for some galaxies from THINGS survey together with
  fits. Blue thick curves show best fits by dark matter distributions of semidegenerate configurations,
  while magenta thin curves show Einasto profile best fits. Galaxy names where semidegenerate profile
  fits rotation curves better than Einasto profile are emphasized,
  and names where profiles are comparable in fit quality are bolded. See digital version for
  colored plots.}\label{Ein+Ferm}
\end{figure*}

The results of fitting are presented in the Table~\ref{fitres} and on
Fig.~\ref{Ein+Ferm}. Due to the scaling laws recalled in Sec.~\ref{solutions}
the fits obtained here for particle mass $m=10$~keV/$c^2$ can be also
transferred to another particle mass by changing the fitted central degeneracy
parameter from $\theta_0^*$ (see Table~\ref{fitres}) according to the relation
\begin{equation}\label{fits}
\theta_0(m)=\theta_0^*+12.52\log\frac{m}{10\text{
keV}/c^2},
\end{equation}
provided that $\theta_0(m)$ is larger than 20 and the influence of the
degenerate core on rotational velocity is negligible in the observed radial
range. From the values of $\theta_0^*$ obtained for the fitting of the sample
listed in Table~\ref{fitres}, and the lower value of $\theta_0(m)$ from which
the scaling laws (\ref{eq:rcSL}--\ref{eq:Mhalo}) are valid, it is possible to
obtain from (\ref{fits}) a preliminary lower limit for the particle mass
$m\gtrsim$ few keV/$c^2$. Nonetheless this limit should not be considered as an
absolute lower limit for the fermion mass of the model because the bound
$\theta_0(m)\geqslant20$ in the formula above is a numerical limit, and no
underlying physics has been specified here for it. The formal way of providing
an absolute lower limit for the particle mass of our model when applied to
typical spiral galaxies has been found in \cite{2013PRL..R}, and yields roughly
an order of magnitude less than the one inferred here.

From the 16 galaxies analyzed, our model has minimum BIC value in 5 cases
(NGC2366, NGC2841, NGC2903, NGC2976, NGC3521), Einasto model in 10 cases
(NGC2403, NGC3031, IC2574, NGC3621, NGC4736, DDO154, NGC5055, NGC6946, NGC7331,
NGC7793), and in the case of NGC3198 Burkert model is the best one, marginally
better than ours, which in turn is marginally better than Einasto. Besides this
general comparison in which apparently Einasto model is preferred against our
model, there is a more relevant comparison which must be made considering that
the semi-degenerate model provides cored halos only. For this, we compare the
Einasto model against the semi-degenerate one for the sub-set of galaxies which
are cored-like (i.e. with Einasto index $n\lesssim4$), and then the same
comparison is made for the sub-set of galaxies which are cuspy-like (i.e. with
Einasto index $n>4$). The important outcome of this new BIC comparison is that
our model is \textit{equivalently} as good as Einasto for the cored-like
sub-sample, that is connected to the fact that Einasto profile with $n\sim1$
provides rotational curves that in a wide range of radii is quite close to the
one of ours at transition from region III to region IV. For cored-like galaxies
in 6 cases our model has lower BIC number than Einasto model (NGC2366, NGC2841,
NGC2903, NGC2976, NGC3198, NGC3521), and inversely in other 6 cases Einasto is
better (NGC3031, IC2574, NGC4736, DDO154, NGC5055, NGC7793). Instead, for the
cuspy-like sub-sample (NGC2403, NGC3621, NGC6946, NGC7331), in \textit{all} the
cases Einasto model has lower BIC numbers as logically one may expect due to
the cored nature of the semi-degenerate halos.

If we take only the models with significant fits, i.e. with reduced $\chi^2$
less than one at least for one fit, then our model is preferred in 3 cases
(NGC2366, NGC2903, NGC2976), Einasto one in 5 cases (NGC2403, IC2574, NGC3621,
DDO154, NGC7331), and Burkert fit for NGC3198 is also significant. It is
remarkable that neither Navarro--Frenk--White model, nor pseudo-ISO model is
preferred against others in the THINGS sample.

If we take into account only two-parametric models, then we have the same 5
aforementioned cases for our model to be the best plus the case of NGC5055, NFW
model is preferred in 2 cases (NGC2403, NGC6946), pseudo-ISO one in 3 cases
(NGC3621, DDO154, NGC7331), and Burkert model in 5 cases (NGC3031, NGC3198,
IC2574, NGC4736, NGC7331). Taking only significant fits, we get the best
performance of our model in 3 cases (NGC2366, NGC2903, NGC2976), NFW in 1 case
of NGC2403, pseudo-ISO in 2 cases (DDO154, NGC7331), and Burkert in 2 cases
(NGC3198, IC2574). From this we can conclude that our model is the best
two-parametric model of the set considered.

It is interesting to make pair comparison of our model with Burkert profile: it
is preferred statistically in 6 cases (NGC2366, NGC2841, NGC2903, NGC2976,
NGC3521, NGC5055) and is disfavored in 10 cases. However, in all these cases
besides IC2574 the preference is only marginal.

It should be mentioned that the velocity profile of the semidegenerate
configuration used for fitting is exact only in the case of Dark Matter
domination and thermal equilibrium at all radii, that can be not the case of
real galaxies. However, it is especially interesting that even such a
simplified model provides good correspondence to empirical rotational curves.

\section{Conclusion}\label{conclusion}

It follows from the results of fitting that the semidegenerate fermionic
distributions can fit dark matter in the THINGS sample of galaxies at least as
well as other profiles considered in the literature, with the important
"revenue" that this profile is theoretically motivated, and is not
phenomenological as most of the others. The cases when Einasto profile fits
rotational curve much better than semidegenerate profile show the general cuspy
behaviour of dark matter distribution, possibly representing a special class of
galaxies that are still not completely relaxed.

While Einasto profile is a pure phenomenological one based on best fit of the
observational and numerical simulation data, our profile is derived from the
first principles and based on the General Relativistic treatment of
self-gravitating neutral fermions. There is the distinct possibility that our
treatment gives the conceptual physical motivation for the existence of the
cored Einasto profile directly from the structure of the microphysical
constituents of dark matter.

\begin{acknowledgements}
The authors wish to thank Prof. Herman Mosquera Cuesta for critical reading of
the manuscript and valuable suggestions.
\end{acknowledgements}



\end{document}